\documentclass{article}

\usepackage{PRIMEarxiv}

\usepackage[utf8]{inputenc} % allow utf-8 input
\usepackage[T1]{fontenc}    % use 8-bit T1 fonts
\usepackage{hyperref}       % hyperlinks
\usepackage{url}            % simple URL typesetting
\usepackage{booktabs}       % professional-quality tables
\usepackage{amsfonts}       % blackboard math symbols
\usepackage{nicefrac}       % compact symbols for 1/2, etc.
\usepackage{microtype}      % microtypography
\usepackage{lipsum}
\usepackage{fancyhdr}       % header
\usepackage{graphicx}       % graphics
\graphicspath{{media/}}     % organize your images and other figures under media/ folder

\title{Framework of a multiscale data-driven DT of the musculoskeletal system
}

\author{
  Martina Paccini, Simone Cammarasana, Giuseppe Patan\'e \\
  CNR - IMATI \\
  Genova\\
  \texttt{\{martina.paccini, simone.cammarasana, giuseppe.patane\}@cnr.it} \\
}

\begin{document}
\maketitle
\begin{abstract}
Musculoskeletal disorders (MSDs) are a leading cause of disability worldwide, requiring advanced diagnostic and therapeutic tools for personalised assessment and treatment. Effective management of MSDs involves the interaction of heterogeneous data sources, making the Digital Twin (DT) paradigm a valuable option. This paper introduces the \emph{Musculoskeletal Digital Twin} (MS-DT), a novel framework that integrates multiscale biomechanical data with computational modelling to create a detailed, patient-specific representation of the musculoskeletal system. By combining motion capture, wearable sensors and medical imaging, the MS-DT enables the analysis of spinal kinematics, posture, and muscle function. An interactive visualisation platform provides clinicians and researchers with an intuitive interface for exploring biomechanical parameters and tracking patient-specific changes. Results demonstrate the effectiveness of MS-DT in extracting precise kinematic and dynamic tissue features. This framework provides high-fidelity modelling and real-time visualisation to improve patient-specific diagnosis and intervention planning.
\end{abstract}
\section{Introduction}
\emph{Musculoskeletal disorders} (MSDs) are among the leading causes of disability worldwide, significantly impacting patients' quality of life and placing a considerable burden on healthcare systems. These disorders comprehend a wide range of conditions, including degenerative diseases (e.g., osteoarthritis, intervertebral disc degeneration), inflammatory disorders (e.g., rheumatoid arthritis), traumatic injuries (e.g., ligament tears, fractures), and spinal deformities (e.g., scoliosis, kyphosis). Chronic MSDs often lead to functional impairments, persistent pain, and reduced mobility. Sometimes these issues can be controlled with conservative treatments, making physiotherapy essential for restoring strength, flexibility, and proper biomechanics. However, when conservative treatments fail to alleviate symptoms, patients may require surgical procedures. Indeed, spinal fusion, joint replacement, or tendon repair could restore function and relieve pain, particularly in progressive degenerative or structural conditions such as severe osteoarthritis or spinal deformities (e.g., scoliosis and kyphosis).

The Digital Twin (DT) paradighm, defined as an adaptive model of a complex physical system~\cite{He2021}, can hep the effective management of most aspects of MSDs. Indeed the application of the DT paradighm to the musculoskeletal system provides a comprehensive and patient-specific representation of the musculoskeletal  health, with various applications. In rehabilitation optimisation they help patients prepare for surgery by assessing muscle strength, joint mobility, and compensatory movement patterns, which are crucial for minimizing surgical risks and improving post-surgical outcomes. In surgical planning they facilitate the minimisation of intraoperative risks. During the follow-up, they support prediction and evaluation in time. 

However, a key challenge in DT is represented by the integration of the various multi-modal data for a comprehensive evaluation of the system. Given the amount of diagnostic and monitoring devices available for the musculoskeletal  system, including wearable sensors, it represents an excellent opportunity for the development of a DT with a multiscale approach based on data integration. Among the most relevant are imaging techniques such as ultrasound (US), magnetic resonance imaging (MRI), and computed tomography (CT), each of which provides distinct insights ranging from anatomical to functional information. Additionally, inertial sensors and motion capture systems contribute to the evaluation of kinematic and kinetic parameters, while wearable technologies like surface electromyography (sEMG) enable the monitoring of physiological and biomechanical metrics~\cite{diniz2025digital}.

For this reason the focus of this work is the integration of musce-skeletal data at multiple level to support the framework of a DT. Thus, this paper introduces an \emph{integrated musculoskeletal Digital Twin} (MS-DT) to support the management of MSDs.

%For preoperative planning, the MS-DT aids spinal deformity correction by generating a patient-specific 3D spinal model through CT and US fusion, facilitating optimized screw placements and fusion level selection while minimizing intraoperative risks. Furthermore, physiotherapy aspects and preoperative optimization can also be managed through the MS-DT, 

%In the postoperative phase, 3D motion analysis  Additionally, physiotherapy plays a crucial role in both non-surgical and post-surgical rehabilitation. The MS-DT supports adaptive rehabilitation programs, dynamically refining physiotherapy based on real-time patient feedback, ensuring optimal recovery and improved functional outcomes. By continuously monitoring muscle adaptation, joint mobility, and movement patterns, the MS-DT can guide clinicians in adjusting exercise regimens and identifying potential complications early in the recovery process.

%The development and validation of this framework require robust performance across varying acquisition conditions and real-time processing capabilities suitable for clinical workflows. Addressing anatomical variability and seamlessly integrating segmentation outputs with other modalities are critical to achieving a truly patient-specific DT model. By bridging multimodal data sources, the MS-DT can improve musculoskeletal healthcare by enhancing diagnostic accuracy, guiding surgical interventions, and optimizing preoperative strategies and postoperative rehabilitation to enhance patient outcomes.

Unlike generalized DTs that aim at considering every aspect of health, or specific DTs that focus on isolated structures, our DT situates itself in the middle ground, providing a comprehensive yet targeted model of the MS system. Moreover, given that different types of data offer insights at varying scales and levels of granularity, it was a natural step to integrate multiple data sources in order to derive insights that would not be accessible through any single source in isolation.
The proposed MS-DT framework introduces the following contributions: 
\begin{itemize}
	\item A multiscale, data-driven digital representation of the subject.
	\item A first level of integration, based on heterogeneous data sources, in which data acquired from various devices are combined to offer clinical practitioners a comprehensive assessment of the patient.
	\item A second level of integration, based on features extracted from the data, where the features are organized into a graph structure representing the individual patient.
	\item Inference performed on the graph-based representation, enabling insights into the patient's clinical trajectory and suggesting potential diagnostic examinations or prescriptions.
\end{itemize}

Globally, the resulting MS-DT supports: (i) precision-driven surgical planning, reducing intraoperative risks by integrating different imagin modalities with a markeless image fusion system; (ii) objective, real-time tracking of recovery, optimising the rehabilitation through the combination of 3D video and motion sensors data integration;(iii) muscle functionality evaluation through sEMG signal analysis (iv) patient-specific treatment orientation support, through multi modal data graph representation and inference improving long-term outcomes.

After an analysis of the state-of the-art (Sect.~\ref{sec:related_works}), the paper presents an overview of the DT framework and the multiscale approach (Sect.~\ref{Sect:OverviewSect}), then explores the actual construction of the MS-DT with detail of the modelling techniques and integration involved (Sect.~\ref{Sect:engines}). Finally we discuss the results obtained and how the framework can easily integrate with existing open-source modelling software for further improvements (Sect~\ref{Sect:discussions}). 

\section{Related work\label{sec:related_works}}
DTs in healthcare represent a transformative concept in continuous evolution aiming to create virtual representations of patients' physiology that evolve dynamically with real-time data. The ambition of DTs in medicine is to provide a predictive and personalised model of an individual’s health status. However, despite the theoretical advancements, there are challenges in data collection, data fusion, and accurate simulation at this stage~\cite{Sun2022}. The construction of a comprehensive DT inherently involves multidisciplinary competencies, including bioengineering, clinical sciences, AI, and data security, which complicates their clinical translation. A particularly critical challenge—central to this work—is the integration of \emph{multimodal data} from heterogeneous sources (e.g., imaging, biomechanics, sEMG, clinical scores). While multimodal data fusion is essential to improve DT accuracy and applicability, it remains hindered by fragmented data formats, non-standardised interfaces, and the lack of universally accepted architectures~\cite{Diniz2025}.
%Scrivere che la costruzione di un DT per quanto si reisca a semplificare, per natura intriseca coinvolge diverse competenze (TROVARE PAPER DA CITARE)
    
\paragraph{Generalized vs. Specific DTs in Healthcare}
Granularity is a critical aspect of DTs in healthcare, defining the level of detail and specificity in the model. Existing DTs in healthcare can be broadly categorized into two types: generalized DTs that aim at a holistic patient representation to model an individual’s entire physiological system  (usually referred to as Digital Human Twin, DHT), and highly specific DTs that focus on particular organs or joint motion simulations. The most ambitious application of DTs aims to achieve a complete digital representation of the physical patient to improve the prevention, diagnosis, and treatment of diseases. However, available applications mostly concern the simulation of single anatomical structures (tissues or organs), physiological processes, or metabolic interactions \cite{Cellina2023, Armeni2022}. Works that present DTs attempting to simulate every aspect of human health with a bidirectional connection with real-time data, face enormous complexities. This results in hypothetical frameworks without realistic applications or in simulation systems that have no actual relation with real-time data acquisition~\cite{Wooley2023}. The key challenges related to the development of DHTs span across multiple domains. Multidisciplinary cooperation is essential, requiring expertise from fields like brain science, psychology, biomechanics, AI, and data analytics to create high-fidelity models. Data acquisition and integration require real-time synchronization and interoperability between diverse sources like wearable devices, medical records, and imaging data. Additionally, data quality, accuracy, and bias issues can compromise DHT reliability, with fragmented, incomplete, or biased datasets leading to flawed outcomes. Multimodal data fusion is crucial for the integration of heterogeneous data sources while filtering redundant information. Autonomy and decision-making require careful consideration to balance automated processes with human oversight, preventing errors from incomplete or biased data. Specifically, multimodal data fusion still lacks robust and validated pipelines capable of combining biomechanical, physiological, and clinical information into unified models~\cite{diniz2025digital}. Finally, computing infrastructure advancements in high-performance computing, 5G, AR/VR, and blockchain will be pivotal in enhancing DHT accuracy, security, and real-time processing capabilities. Addressing these challenges is crucial to unlocking the full potential of DHTs while ensuring ethical, secure, and reliable applications~\cite{Katsoulakis2024, He2024}.

To reduce the influence of these challenges several works present DT highly focused on a specific organ, simulation or pathology. Considering the musculoskeletal  system the DTs proposed in the literature are typically highly focused on a joint, organ or specific task. The DT in ~\cite{Chen2024} considers the whole body but focuses on a specific goal, i.e., 3D pose reconstruction utilising monocular camera videos as input. Other DTs have been developed for the production of simulated data for future AI network training: electromyography signals~\cite{Maksymenko2023}, and gait analysis~\cite{Uhlenberg2023} algorithm testing.In ~\cite{Hernigou2021}, leveraged artificial intelligence, machine learning technology and DT paradigm to identify a real personalized motion axis of the tibiotalar joint. 3D models of distal extremities were generated using computed tomography data of normal patients. DTs were used to reproduce the mobility of the ankles, and the real ankle of the patients was matched to the DT with machine learning technology. Main examples include a DT simulating the vertebroplasty procedure and its impact on mechanical stability in cancer patients, an integrated DT for lumbar spine analysis~\cite{Ahmadian2022}. A \emph{finite element model} (FEM) of the lumbar spine was developed using CT and constrained by the body movements, which were calculated by the inverse kinematics algorithm. The Gaussian process regression was leveraged to train the predicted results and create the DT of the lumbar spine in real time. Finally, a 3D virtual reality system was developed using Unity3D to display and record the real-time biomechanics performance of the lumbar spine during body movement. This DT has been tested and applied to real data. Still focusing on spine in~\cite{Sun2023} a DT has been generalised to the personalised diagnostic and treatment of musculoskeletal system diseases, but at the current state, it remains a theoretical framework.

These specific DTs, such as those modelling knee joint dynamics or cardiac function, provide highly detailed simulations or focus on a specific pathology, but lack the systemic interactions necessary for comprehensive patient modelling and understanding. The challenge lies in achieving a balance between generalization and specificity, ensuring that the DTs remain computationally feasible while retaining high fidelity in their predictive and diagnostic capabilities. This gradation in complexity necessitates a multiscale approach, integrating data from molecular, cellular, tissue, organ, and whole-body levels.

\section{Framework overview\label{Sect:OverviewSect}}
The MS-DT framework follows a multiscale approach with a modular structure, designed for the integration and correlation of data across scales.
\begin{figure}[t]
 \centering
\includegraphics[width =1 \linewidth]{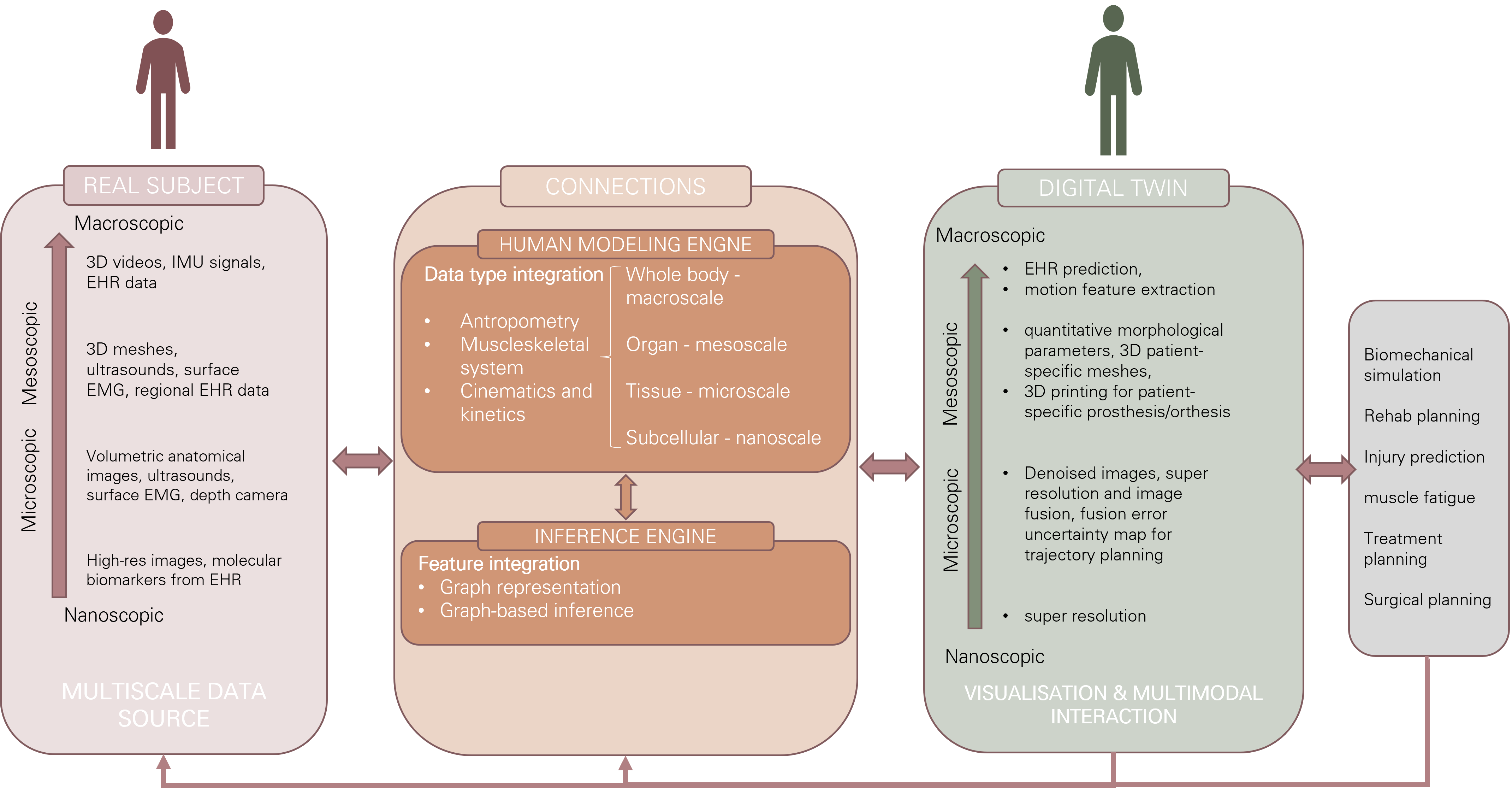}
	\caption{MS-DT framework overview and main component organisation\label{fig:FrameworkOverview}.}
\end{figure}

The DT framework (Fig.~\ref{fig:FrameworkOverview}) consists of three main sections: \textit{Multiscale Real Subject}, which manages the data acquired and recorded from the patient at different scale levels; \textit{Connections}, which communicate bidirectional with input data and the virtual representation to provide relevant information; and \textit{Multiscale DT}, which provides an interactive environment where physician and surgeons can visually and quantitatively evaluate the patient's situation. The final element of the MS-DT framework (grey block in Fig.~\ref{fig:FrameworkOverview}) is represented by state-of-the-art methods and available open-source tools that can interact with the DT for further evaluations and could be connected with the other sections (i.e., Real subject and Connections) to provide feedback to the already included models (we refer the reader to Sect.~\ref{Sect:discussions}). 

\paragraph{Multiscale Real Subject\label{Sect:MSdata}}
Creating a multiscale framework for a twin, particularly for the musculoskeletal system, involves defining hierarchical levels of organization and associating data with each scale. From the data our MS-DT aims at interpret and infer key aspects of MSDs and their surgical or physiotherapy-based treatment. The anatomical granularity of the data is the key to the hierarchical structure, going from nanoscopic to macroscopic scale. The organisation of the input data is reflected in the virtual description of the subject (i.e., the output of the MS-DT).

The \emph{Macroscale} analyses the overall body mechanics and dynamics, movement patterns, and system-level interactions. The \emph{Mesoscale} focuses on functional and anatomical relationships within specific regions, such as joints, muscles, and bones. At the \emph{Microscale}, the focus shifts to the structural and functional properties of individual tissues, including muscles, tendons, bones, and nerves. Finally, the \emph{Subcellular or Nanoscale} level is dedicated to understanding molecular dynamics, cellular processes, and localized physiological phenomena within tissues. 

By integrating these different scales of analysis, MSDs management can be optimized through a comprehensive, multilevel approach, improving both surgical planning and rehabilitation strategies adapted to the individual patient needs.

%\emph{Inertial Measurement Unit} (IMU) sensors aid in assessing body kinematics by measuring the acceleration and angular velocity of the considered segments, with possible continuous monitoring. To further personalise the evaluations at this scale, \emph{Electronic Health Records} (EHR) provide insights into medical history, physical activity levels, and general functional assessments. 
%Data sources at this level include volumetric anatomical imaging techniques such as MRI and CT for detailed tissue characterization, real-time ultrasound for tracking dynamic tissue deformation, and surface EMG for assessing electrophysiological activity at the muscle-tissue level.

\paragraph{Connections module\label{Sect:Connections}}
Following the naming proposed by~\cite{He2024}, the Connections section acts as the core computational layer, linking real-world patient data with the virtual model through two key components.

The \emph{Human Modelling Engine} (HME) module constructs a biomechanical model of the patient by integrating anthropometric data, musculoskeletal system modelling, and kinematics/kinetics analysis. It processes information across the four levels: whole-body (macroscale), regional (mesoscale), tissue (microscale), and subcellular (nanoscale). This ensures a multiscale representation of movement, forces, and functional anatomy. To this aim, the HME module integrates the different types of data to provide the information required at each scale, which are not obtainable for the single data sources separately.

The \emph{Inference Engine} performs a second level of integration by analysing the features extracted from multimodal data. It constructs a graph-based representation of the individual patient, where clinical, biomechanical, and physiological features are organized into a structured model. Based on this graph, the engine performs inference to assess the patient's clinical trajectory and to suggest potential diagnostic examinations or treatment prescriptions.Moreover, the system continuously updates the DT Representation through bidirectional interaction with the Multiscale Data Source, refining its insights with new data. This dynamic, patient-specific model supports informed clinical decision-making, personalized treatment planning, and proactive rehabilitation monitoring.

\paragraph{Multiscale DT\label{Sect:DTrapVis}}
On the one hand the DT generated within this framework is a virtual counterpart of the patient, continuously updated with real-time and historical data to provide a dynamic and interactive visualisation of biomechanical function. The DT operates at multiple scales, reflecting macro-, meso-, micro-, and nanoscale characteristics of the musculoskeletal system and the input data.
On the other hand the DT consists also of a global representation based on the feature extracted from the integration of the various data sources, which has the goal of supporting and guiding the patient's path and the physician's decision making.

\begin{table}[!h]
\begin{center}
\caption{Overview of the heterogeneous data source handled by the MS-DT in relation to the multiscale approach\label{tab:dataAcquisition}}
	\begin{tabular}{|p{3cm}||p{3cm}|p{3cm}|p{3cm}|}
		\hline
		\textbf{Scale} & \textbf{Focus}&\textbf{Data types} \\
		\hline\hline
		Macroscale & Whole-body mechanics & 3D videos, IMU signals, EHR data \\ \hline		Mesoscale & Regional relationships & 3D meshes, US, sEMG, regional EHR data \\ \hline
		Microscale & Tissue and sub-tissue dynamics  & Volumetric anatomical images, US, sEMG, depth camera acquisitions \\ \hline
		Subcellular/Nanoscale & Cellular and molecula processes & High resolution images, molecular biomarkers from EHR \\ \hline
	\end{tabular}
\end{center}
\end{table}

\section{DT development\label{Sect:engines}}

\subsection{Data acquisition}
The acquisition of \textit{multimodal biomechanical and neuromuscular data} is fundamental for constructing the MS-DT. Table~\ref{tab:dataAcquisition} resumes the different heterogeneous data considered and processed in our MS-DT. Given the focus of our framework on data integration, a single data typology, (e.g. US imaging) can be leveraged for more than one scale by changing its analysis, as well as a specific processing (e.g., super-resolution) can be applied to different imaging sources at different scales.  To achieve high-resolution, real-time motion tracking and muscle activity assessment, we integrate wireless \emph{Inertial Measurement Unit} IMUs and sEMG. SEMG sensors detect muscle activation levels while IMU captures linear acceleration, angular velocity, and spatial orientation. These non-invasive, body-worn sensors provide continuous kinematic and muscular feedback.

In our study, we utilized Cometa Srl's PicoX sensors, which integrate both IMU and sEMG capabilities. These sensors provide high-fidelity motion and muscle activation data, ensuring precise monitoring of biomechanical function. The PicoX IMUs offer 9 degrees of freedom (DoF) with data fusion capabilities, capturing accelerometer, gyroscope, and magnetometer signals in real time. Additionally, the integrated sEMG system records muscle activity with a sampling frequency of 2000 Hz, ensuring high-resolution signal acquisition for accurate neuromuscular analysis.

Positioning sEMG and IMU sensors for muscle activity evaluation follows standardised protocols. Regarding the evaluation of the lumbar spine, sEMG signals are recorded from the erector spinae muscles during the execution of different movements. The electrodes are positioned bilaterally at the L4 and L2 vertebral levels, with EMG1 and EMG2 sensors placed laterally at L4, and EMG3 and EMG4 sensors at L2~\cite{ekstrom2020model}. This configuration allows us to evaluate the activation of the paraspinal muscles, which is essential for quantifying spinal stability or spinal curve progression. 

Beyond the lumbar spine, the PicoX sensors can be applied to analyse kinematics and muscle activation patterns across the entire body. By following appropriate protocols, these sensors allow for comprehensive motion analysis, including upper and lower limb assessments, postural control studies, and gait evaluations. The adaptability of the system makes it a versatile tool for assessing a wide range of musculoskeletal conditions and rehabilitation progress.

Additionally, the same movements performed during IMU and sEMG measurements can be simultaneously recorded using 3D video acquisition systems such as depth cameras. This enables further kinematic analysis by extracting movement trajectories, joint angles, and overall biomechanical parameters. The integration of 3D video data with sensor-based acquisition enhances the accuracy and depth of motion analysis, allowing for a more detailed evaluation of movement disorders, rehabilitation progress, and performance optimization. Indeed, 3D videos may allow the analysis of kinematic features in parts of the body where IMU sensors are not placed. 

For a dynamic assessment of tissue behaviour, US imaging plays a fundamental role, as it allows for the evaluation of soft tissue structures during movement execution~\cite{tozzi2011fascial, bubnov2018ab1199}. With the diffusion of portable echography, such as hand-handled or ultraportable probes, the US imaging technique is highly suitable for DT development. For this study we considered both images from standard ecographs and portable US probe, such as the Philips Lumify\texttrademark . While CT and MRI provide highly accurate, less noisy, and anatomically detailed data, US enables real-time tissue analysis, capturing muscle contractions, tendon deformations, and joint mechanics in action. However, despite its advantages in dynamic assessment, US data can be noisier and more challenging to interpret due to variations in probe positioning, operator dependency, and lower spatial resolution. Integrating US with CT, MRI, and motion capture data enhances the capability of the musculoskeletal DT, providing both static anatomical precision and real-time functional insights. 

Transversal to all the imaging modalities integrated within the MS-DT framework, such as ultrasound, MRI, CT, we incorporate real-time denoising and super-resolution techniques to enhance image quality and diagnostic value. These processes are implemented using deep learning models that are adaptable to different data types and anatomical regions. Real-time denoising reduces acquisition noise while preserving fine anatomical details, facilitating clearer interpretation and more reliable feature extraction across imaging sources~\cite{cammarasana2022real}. Similarly, real-time super-resolution enhances both spatial and temporal resolution, allowing for the extraction of detailed structural and functional information in an interactive and clinically usable form~\cite{cammarasana2023super}. By applying these enhancement methods consistently across modalities, we ensure high-quality data integration and improved performance of subsequent processing steps such as segmentation, registration, and feature analysis.

\emph{Electronic Health Records} (EHR) include a complete set of patient data: personal details, medical histories, images, clinical reports, pathologies and test results, thus they can serve at multiple scale levels: from Nanoscale to Macroscale. Artificial intelligence tools applied to EHR data can assist with tasks such as analysing disease-symptom relations, diagnostic inference, and improving healthcare decision-making. The analysis of drug administration supports the personalization and precision of the therapy to patients, reduces the risk of ineffective treatments and adverse drug reactions, and improves drug efficiency and management. In literature, \emph{Graph Convolutional Networks} (GCNs) are applied to predict patients’ interventions in ICU (e.g., mechanical ventilation interventions)~\cite{xu2024predicting}, patients’ readmission by analysing free text data of EHR~\cite{lu2021predicting}, and drug-disease interaction against COVID-19~\cite{che2021knowledge}. 

\subsection{Human modeling engine\label{Sect:HME}}
The \emph{Human Modelling Engine} (HME) serves as a robust platform for understanding complex human biomechanics and anatomical structures through various data acquisition techniques and analytical methods. It integrates heterogeneous data sources and types to model the musculoskeletal system. In the following we will describe the implementation of data type integration in our MS-DT.

\paragraph{3D digital representation of the patient}
\begin{figure}[t]
	\centering
	\includegraphics[width =0.6 \linewidth]{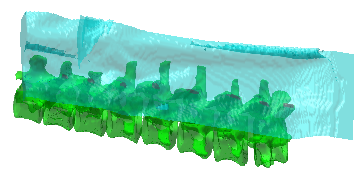}
	\caption{3D visualisation of vertebral landmarks in 3D space with respect to the skin of the subject\label{fig:skinlandmarks}.}
\end{figure}
To obtain the digital multi-scale representation of the musculoskeletal  system of each subject we leverage 3D surface models of the body and bones obtained from the respective segmentation in volumetric images such as MRIs or CT. The skin segmentation for the 3D body representation is obtained automatically from both volumetric image sources, through a \textit{graphics-based approach}~\cite{paccini2024us}. While the segmentation of the patient's bones, especially of the vertebral spine, are obtained from CT images knowing the respective HU value or though segmentation tools such as the Totalsegmentator~\cite{wasserthal2023totalsegmentator}.

This 3D shape model representation plays a crucial role in visualising both the DT and the results derived from data analysis. By incorporating both the skin surface and the vertebral spine into a patient-specific 3D model, it enables more accurate identification of anatomical landmarks, such as the vertebral transverse processes, which in turn provides critical insights into spinal alignment and curvature (Fig.~\ref{fig:skinlandmarks}). Furthermore, these 3D models of the skin and vertebrae form the foundation for various data integration strategies and subsequent analyses that contribute to the comprehensive development of the DT.
\begin{figure}[t]
\centering
\includegraphics[width =0.7 \linewidth]{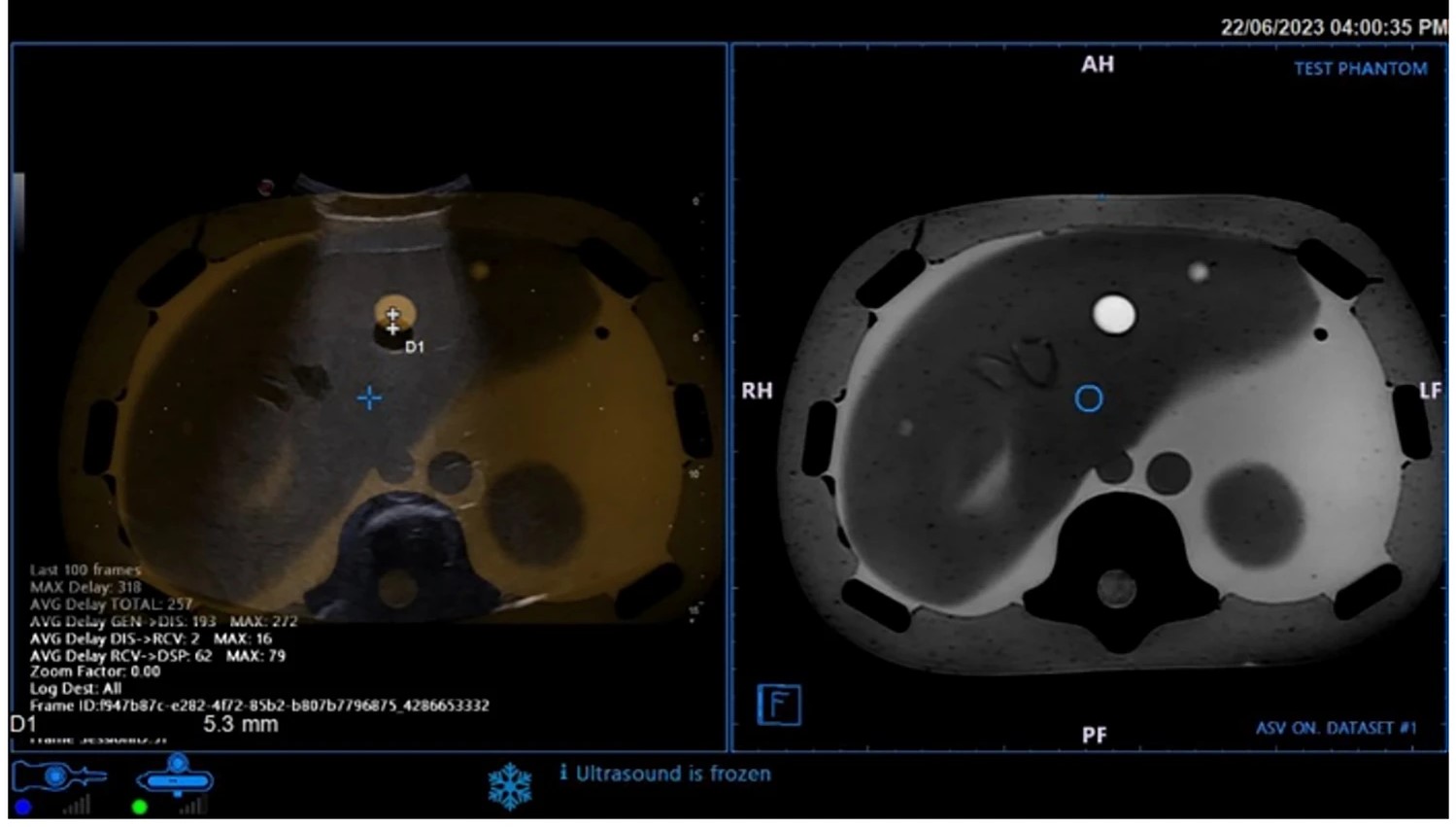}
	\caption{Image fusion implementation, results show millimetre accuracy even with symmetrical anatomical regions and without placing physical markers. Image courtesy of Esaote S.p.a\label{fig:imageFusion}.}
\end{figure}

\paragraph{3D surfaces and volumetric images integration for fusion imaging}
Image fusion enables the integrated visualisation of heterogeneous imaging modalities, providing a comprehensive and enriched information set for patient-specific assessment. To support advanced clinical workflows, particularly in preoperative and intraoperative planning, we integrate into the MS-DT (Multiscale DT) an MR–US fusion system that enables real-time visualisation without relying on physical markers or AI model training, and is adaptable to various anatomical regions~\cite{paccini2024us}. This system requires the integration of data from different sources: a portable 3D camera for surface acquisition, an electromagnetic tracking system, US imaging components and 3D anatomical images (CT/MRI). As, metioned earlier, the key of the integration resides in the 3D surface representation of the patient's skin both from the 3D camera acquisition and from the anatomical image segmentation. The fusion algorithm comprises two main phases: skin segmentation and rigid co-registration. The co-registration step aligns the 3D skin surface extracted from MR images with the camera-acquired patient surface, enabling accurate, rapid fusion that supports both surgical planning and intraoperative navigation. The co-registration accuracy of US and MRI was assessed using phantoms and volunteers, with a \emph{target registration error} (TRE) ranging from 4.3 mm to 13 mm in phantoms and an average error of 7.4–9 mm in volunteers. The robustness of the co-registration process was further confirmed under varying acquisition distances, camera tilting angles, and virtual landmark displacements, ensuring reliability across different clinical scenarios~\cite{paccini2024us}. Compared to existing methods, this image fusion system is suitable for MS-DT application since it does not require physical landmark placement, reinforcing its potential for real-time patient monitoring, preoperative planning, and intraoperative guidance in musculoskeletal interventions (Fig.~\ref{fig:imageFusion}).
The visualisation of both US and MRI/CT images, either superimposed or displayed in parallel, is essential for a comprehensive assessment of the patient's musculoskeletal  functional and anatomical status across all stages of care.
\begin{figure}[t]
	\centering
	\centering
	\begin{tabular}{ccc}
	(a)\includegraphics[height=0.4\linewidth]{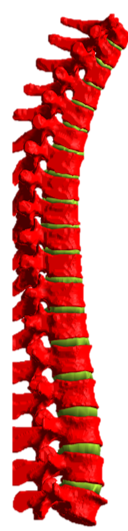}&
	(b)\includegraphics[height=0.4\linewidth]{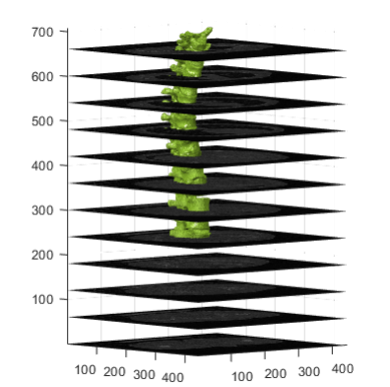}&
		\includegraphics[height=0.4\linewidth]{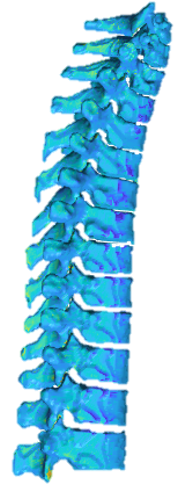}
	\end{tabular}
	\caption{\label{fig:SurfVolIntegration}3D surfaces and volumetric images integration for spine characterisation:(a) extraction of intervertebral space 3D model and (b) integration of volumetric image information on the 3D surface model through texture.}
\end{figure}
\paragraph{3D surfaces and volumetric images integration for anatomical distric characterisation}
The proposed MS-DT integrates multiple data types to enable comprehensive geometrical and tissue-level analysis. Central to this framework is a detailed 3D surface representation of the vertebral spine, which supports the extraction of quantitative geometric parameters crucial for patient-specific assessment (e.g., spinal curvature as measured by the spinal curvature Cobb angle, disc degeneration, and alignment). In contrast to conventional approaches that focus only on adjacent segments, our MS-DT enables the evaluation of the entire spinal column. The 3D spine model allows for the extraction of intervertebral spaces and their spatial distribution, providing insights into how these structural relationships influence the patient’s overall posture.

To enrich the 3D representation, we integrate anatomical imaging at the surface level using a texture mapping algorithm~\cite{paccini2020analysis}. This algorithm maps grey-level values from the CT/MRI images onto the corresponding regions of the 3D bone surface, enabling a fused representation where geometric features and tissue conditions can be evaluated simultaneously. Beyond surface-level assessment, tissue characterization is extended into the interior of the vertebral bodies by extracting grey-level distributions within a 3D sphere centered at each vertebra’s centroid. This approach supports the evaluation of internal bone quality and tissue integrity~\cite{paccini20233d}. Indeed, the combined evaluation of geometrical parameters and tissue composition using HU values reveals clear distinctions between healthy and degenerated bone structures, supporting precise diagnostics.

By combining these modalities, the MS-DT enables the identification of clinically relevant pathologies such as vertebral fractures, osteoporotic changes, and disc degeneration (Fig.~\ref{fig:SurfVolIntegration}). The ability to visualise both the textured 3D model and the underlying volumetric images provides added value during both preoperative planning and postoperative follow-up. When used in conjunction with multimodal image fusion (e.g., US/MRI), this integrated approach allows for the extraction of features and diagnostic insights that are not achievable when data sources are evaluated singularly.

\paragraph{Integration of 3D camera videos with IMU sensors an sEmg}
The integration of 3D video analysis with sEMG and IMU data within the proposed MS-DT framework enables a comprehensive and physiologically grounded evaluation of patient movement, providing insights that would not be obtainable from isolated data sources. In spinal surgery and rehabilitation contexts, 3D videos can be leveraged in preoperative and postoperative evaluations by allowing the assessment of full-body movement mechanics. Preoperatively, they help clinicians understand the patient’s baseline mobility and plan interventions based on biomechanical behaviour. Postoperatively, they allow for monitoring of recovery progress, early detection of abnormal movement patterns, and tailoring of rehabilitation programs to restore functional motor control~\cite{cammarasana2023spatio}.

The 3D movement data are analysed using both geometric and kinematic descriptors, such as centroid trajectories, point-wise speed, volume changes, and activated voxels, computed over time. Speed histograms, for example, are used to classify body regions into dynamic or static behaviours, supporting action recognition and inter-frame comparison. These kinematic features are further enhanced with shape-based descriptors (e.g., shape context, intrinsic shape signatures), which increase accuracy even when the subject's posture varies significantly across sequences. This analysis helps quantify how specific regions of the body contribute to or compensate for localized dysfunctions. The analysis of 3D videos is applied to different data sets (e.g., point clouds, skeletons), subjects (e.g., man/woman, children), data properties (e.g., noisy point cloud, low/high temporal frequency), both with and without physical markers. We identify similar poses (e.g., squat position) or similar behaviours (e.g., a subject moving toward an object) between two different subjects performing actions. The kinematic and geometric descriptors are easy to compute and allow users to compare different actions without the need for a large data set and a training phase, through a fully unsupervised method; in addition, they achieve good action classification and recognition results, with low computational cost~\cite{cammarasana2023spatio}.

To complement the mechanical understanding of movement, sEMG signals are analysed in both the temporal and frequency domains. Metrics such as \emph{Integrated Electromyography} (IEMG) and \emph{Root Mean Square} (RMS) inform on muscle activation intensity and neuromuscular effort, while \emph{Median Frequency} (MF) and \emph{Mean Power Frequency} (MPF) reveal signs of fatigue and fiber-type recruitment~\cite{Suo2024}. Importantly, this allows for the identification of muscle dysfunctions that are not always apparent through motion data alone. For example, high effort with minimal movement could signal underlying neuromuscular inefficiency or compensation.

This combined evaluation, linking joint kinematics with muscular fatigue and control, enables clinicians to better understand the physiological and anatomical contributors to movement disorders. It helps discern whether observed anomalies stem primarily from mechanical limitations or neuromuscular factors, supporting targeted interventions. The approach is particularly useful for comparing a patient's condition before and after rehabilitation, offering objective measures to assess recovery and treatment efficacy.

Additionally, the inclusion of IMU data, provided by the Cometa system alongside sEMG, offers a further source of kinematic information. This supports validation of the 3D video analysis, ensuring consistency across different modalities. A significant advantage of including 3D video is its ability to capture full-body motion, providing insights into how localized impairments influence overall posture or gait. For instance, dysfunction in a spinal segment can lead to compensatory movements in other regions, and this systemic interaction is clearly visualised through 3D recordings.

\subsection{Inference engine}
\begin{figure}[t]
	\centering
	\includegraphics[width =0.8 \linewidth]{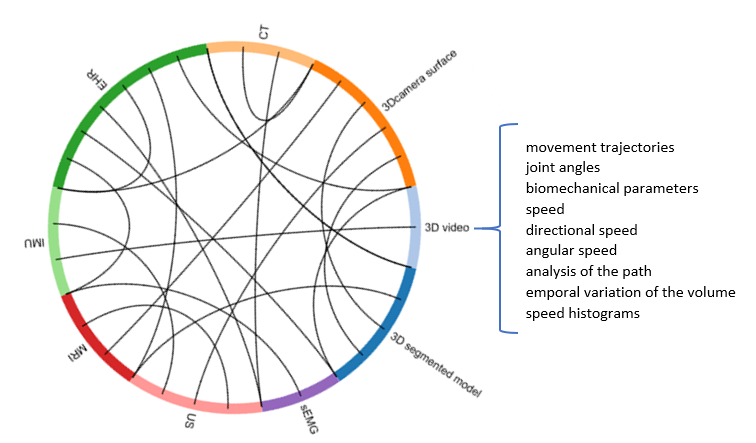}
	\caption{\label{fig:graph}Chord graph representing the single patient.}
\end{figure}
Within the architecture of our MS-DT, the inference engine plays a central role in enabling real-time, patient-specific predictions and decision support. Indeed, it performs the second level of interaction based on data feature. Each data acquisition system mentioned in Sect.~\ref{Sect:HME} is mapped to a feature space, capturing relevant physiological, biomechanical, or functional characteristics of the patient (Fig.~\ref{fig:graph}).These temporally synchronized feature spaces serve as the substrate for the inference engine, which transforms raw data into clinically actionable insights. Indeed, trends emerging across imaging and sensor-derived data can be integrated to estimate the likelihood of disease progression, informing decisions between surgical intervention and conservative rehabilitation strategies.

A central focus of the inferential process is the estimation of the risk of surgical intervention within a 6–12 month horizon in patients affected by degenerative spinal conditions. Structural imaging data (e.g., MRI and CT) provide quantitative indicators such as disc height reduction and facet joint degeneration. Concurrently, biomechanical data from 3D video and IMU systems capture compensatory movement patterns and kinematic asymmetries that reflect functional instability. Complementary information from sEMG reveals muscular fatigue signatures that may indicate chronic overload or maladaptive motor strategies. The inference engine integrates these multimodal features to generate a probabilistic risk score, supporting longitudinal monitoring and clinical decision-making.

This inferential pipeline, from multimodal input to interpretable features and risk stratification, is visually structured through a chord graph representation, which maps the relationships between data modalities, derived features, and predicted outcomes. This graphical abstraction not only supports transparent exploration of the data-feature space but also enhances model interpretability for clinical users.

Importantly, the inference process is not static; it evolves over time as the DT continuously assimilates new patient-specific information. In this dynamic context, the inference engine functions as the computational core of the MS-DT, transforming evolving multimodal representations into predictive models that support personalized diagnosis and treatment planning.
%We propose a graph representation learning model for predicting drug administration using the MIMIC-III database~\cite{johnson2016mimic}, which contains over 53,000 intensive care unit (ICU) admissions data. To this end, we generate a heterogeneous weighted multi-parameter directed graph for the representation of the EHR data of the MIMIC-III database accounting for personal, demographic, diagnostic, and therapeutic features, the proposal of a novel learning-based approach, and the related design and training of a graph convolutional network to predict the drug administration to patients, whose accuracy is around~$75\%$.

\subsection{Visualisation and multimodal interaction}
\begin{figure}[t]
 \centering
\includegraphics[width =1 \linewidth]{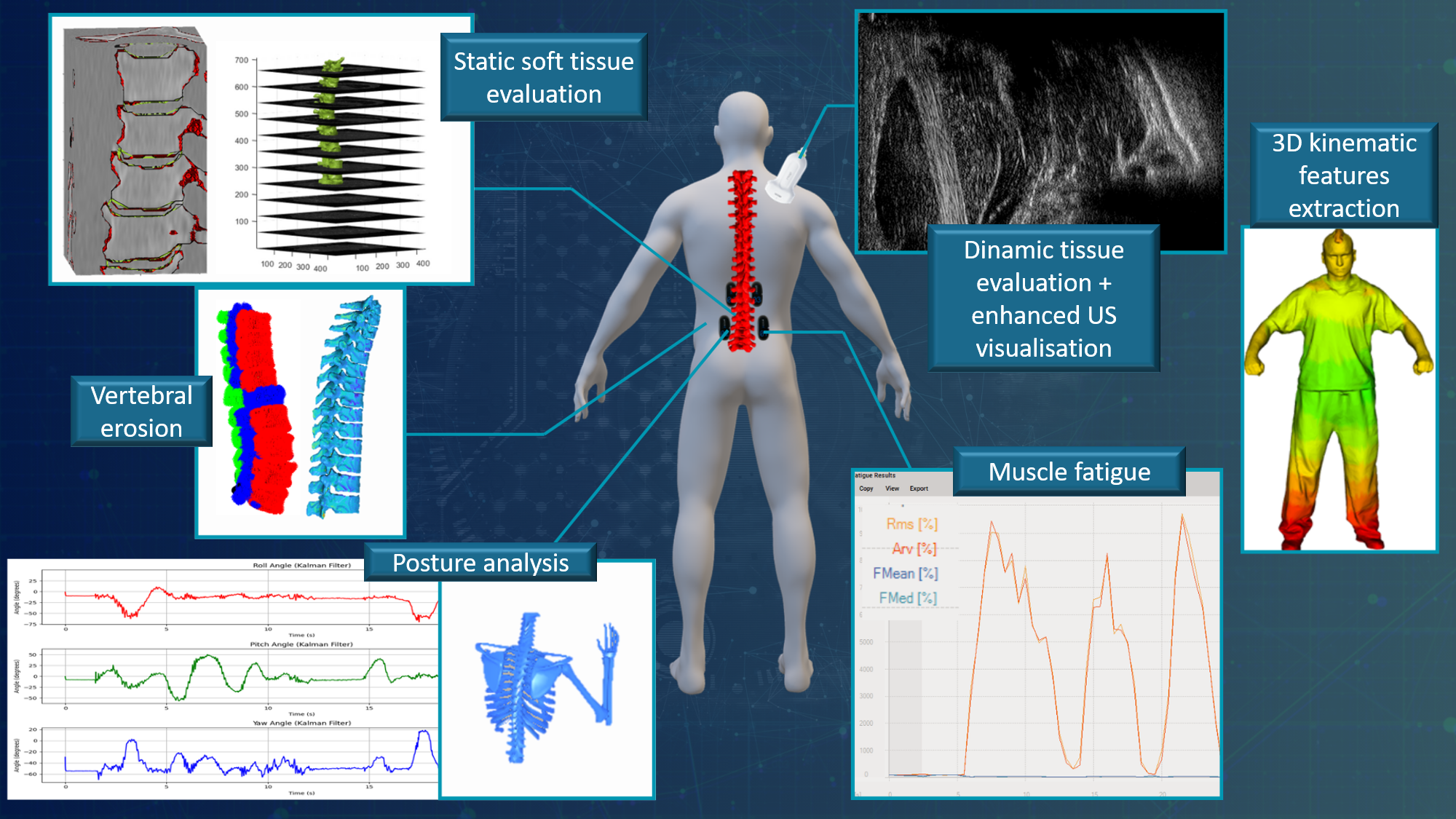}
	\caption{MS-DT visualisation interface allows to consider all the available data of the subject and the results obtained from all the evaluation of the HME.\label{fig:InteractiveInterface}.}
\end{figure}

The interactive visualisation (Fig.~\ref{fig:InteractiveInterface}) provides a comprehensive and patient-specific representation of biomechanical and physiological data, integrating multiple layers of information extracted from the multiscale data sources. This visualisation is directly linked to the Human Modelling Engine, which processes data across different scales, from macroscopic motion features to microscopic and nanoscopic tissue properties, generating an accurate DT of the patient. By leveraging real-time data processing results and inference engine analytics, the visualisation platform enables dynamic exploration of various biomechanical parameters. %including static and dynamic soft tissue evaluation through volumetric anatomical imaging and enhanced ultrasound visualisation, vertebral degradation and erosion analysis for assessing spinal pathologies, and 3D kinematic feature extraction for motion tracking and posture assessment. 

The visualisation interface combines high-resolution 3D models, quantitative metrics, and multimodal data analytics to offer an interactive and layered representation of patient-specific anatomical structures, motion patterns, and functional impairments. Through a bidirectional connection with the multiscale data input, the system ensures real-time updates, dynamically reflecting changes in patient assessments. This interactive visualisation functions not only as a diagnostic aid but also as a decision-support system for clinicians and researchers, facilitating the management of MSDs at different stages by integrating patient-specific morphological and functional parameters. Furthermore, the system incorporates advanced computational techniques such as super-resolution imaging, image fusion, and uncertainty quantification, enhancing accuracy in trajectory planning and enabling error-aware assessments. By bridging data-driven insights with clinical applications, the visualisation platform provides an advanced tool for improving diagnostic precision, optimizing rehabilitation strategies, and supporting preoperative and intraoperative decision-making in musculoskeletal healthcare.
\begin{figure}[t]
 \centering
\includegraphics[width =1 \linewidth]{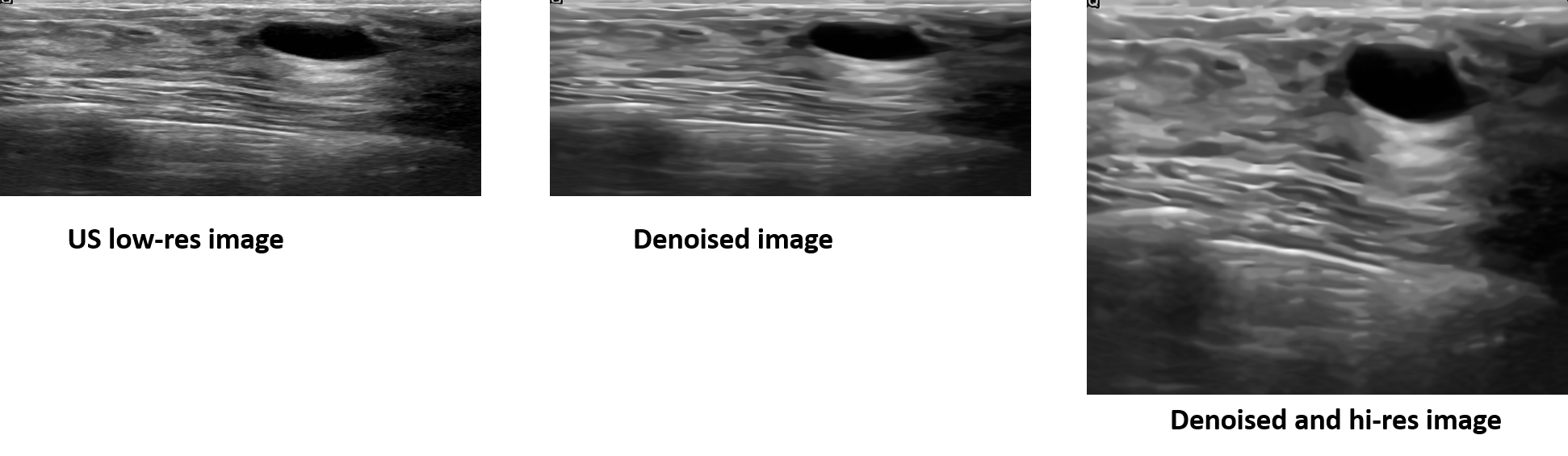}
	\caption{Super resolution and denoising on musculoskeletal  US.\label{fig:superresolution}.}
\end{figure}
%

%\begin{itemize}
    %\item Presentation of system interface with various component results. - FIGURA 3 MA CON PERSONA REALE
%\end{itemize}

\section{Discussion\label{Sect:discussions}}
Implementing the MS-DT framework in clinical and research settings, particularly for spine intervention monitoring, holds transformative potential. It enables personalized assessments of spinal biomechanics, facilitating tailored surgical planning and postoperative rehabilitation strategies. Additionally, the framework supports continuous monitoring of patient progress, allowing for real-time adjustments to treatment protocols. In research contexts, the MS-DT can serve as a valuable tool for studying the biomechanical implications of various spine pathologies and interventions, thereby advancing our understanding and management of spinal disorders.

All the algorithms and modelling methods integrated into the HME of the MS-DT have been designed to be as general and acquisition-agnostic as possible, ensuring seamless adaptability across different imaging modalities and clinical applications. The segmentation and geometric modelling approaches operate independently of specific acquisition techniques, allowing consistent processing of data from MRI, CT, PET, ultrasound, and motion capture systems. Similarly, the image processing components, including real-time denoising, super-resolution, and multimodal fusion, enhance data quality regardless of the source, preserving anatomical and functional details while reducing modality-specific artefacts. The co-registration framework ensures robust alignment of patient-specific data, remaining stable across variations in acquisition parameters, patient positioning, and imaging system specifications. Furthermore, the biomechanical analysis relies on anatomical and functional principles rather than modality-dependent characteristics, ensuring consistent evaluation of movement patterns, soft tissue properties, and musculoskeletal health. By maintaining a high level of generalisation, the MS-DT framework remains highly scalable, facilitating integration into a wide range of clinical applications, data types and features. This modular and adaptable design enables continuous expansion, making the MS-DT a versatile and future-proof tool for personalized medicine, real-time patient monitoring, surgical planning, and therapy optimization.

Moreover, even if part of the data in input has not yet been tested (e.g., microscopic imaging) in the framework, we already considered them as included since the MS-DT has been thought to facilitate the scalability of the overall system. However, given the generality of the HME components some of the evaluation could be directly applied to such data (e.g, denoising and super-resolution)

As current limitations, the integration of complex frameworks, such as the MS-DT, into clinical workflows presents challenges, including the need for standardisation to ensure efficient DT functioning. Moreover, the accuracy of simulations is contingent upon the quality and completeness of input data, which may vary across clinical settings. Interoperability issues with existing healthcare systems and the requirement for continuous updates to maintain the DT's fidelity further complicate implementation. Addressing these challenges is crucial for the successful adoption and efficacy of the MS-DT framework in enhancing patient outcomes. 

Integrating existing biomechanical analysis tools, such as OpenSim~\cite{delp2007opensim} and Mokka~\cite{barre2014biomechanical}, into our MS-DT framework offers significant potential for enhancing data analysis and simulation capabilities. OpenSim provides an extensible platform for developing musculoskeletal models and simulating movement dynamics, enabling detailed assessments of muscle function and joint mechanics. Mokka facilitates the visualisation and analysis of biomechanical data, including markers' trajectories and force platforms, supporting comprehensive evaluations of motion patterns. Moreover, tools such as FEBio~\cite{finley2018febio} and 3D Slicer~\cite{pieper20043d} offer valuable functionalities. FEBio specializes in finite element analysis tailored for biomechanics and bioengineering, enabling the simulation of complex musculoskeletal structures under various conditions. 3D Slicer provides advanced image analysis and scientific visualisation, supporting the processing and interpretation of medical imaging data. By incorporating these tools, our MS-DT framework can achieve more precise modelling and simulation.
\begin{figure}[t]
	\centering
	\centering
	\begin{tabular}{cccc}
		\includegraphics[height=0.2\linewidth]{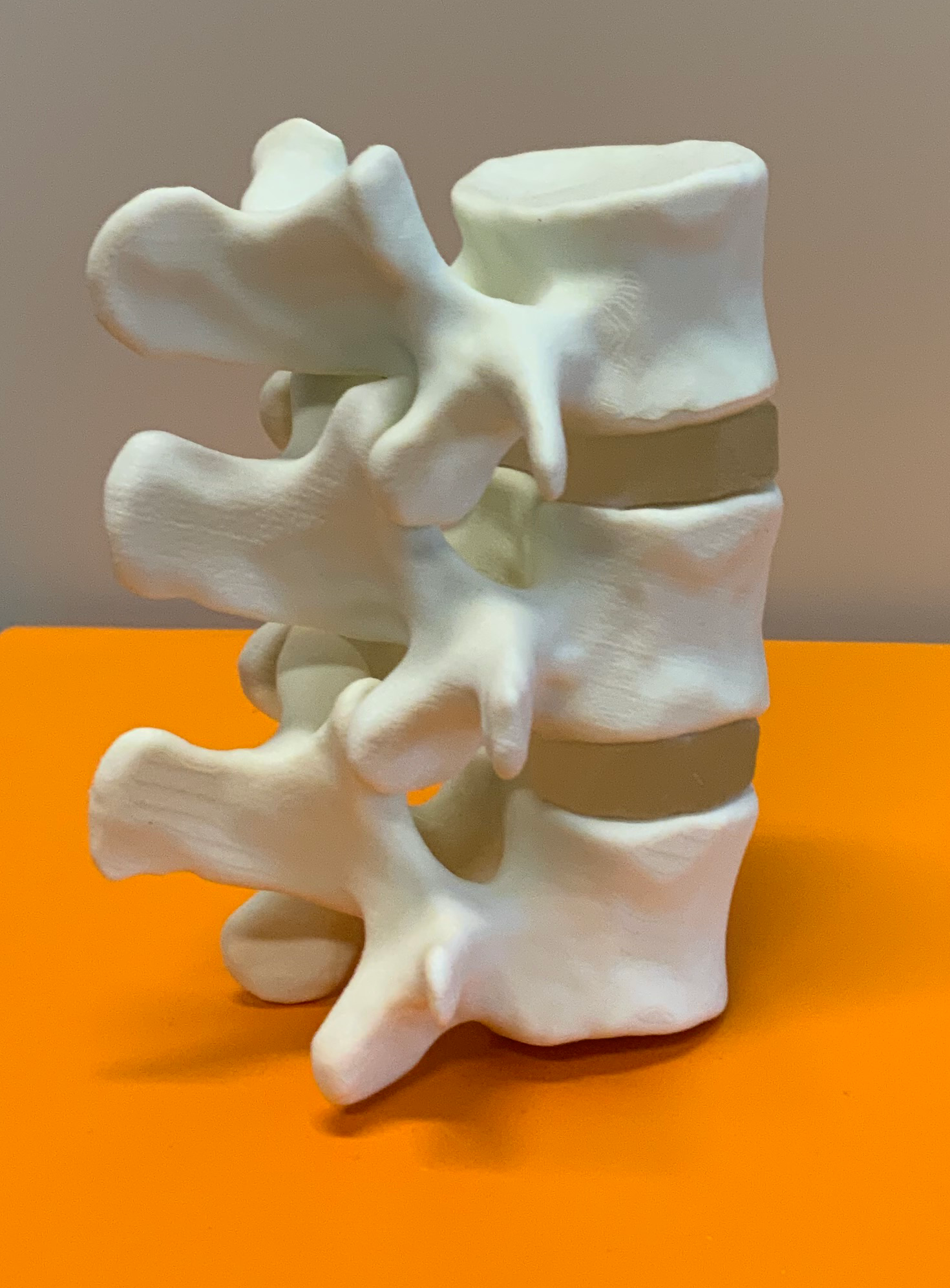}&
        \includegraphics[height=0.2\linewidth]{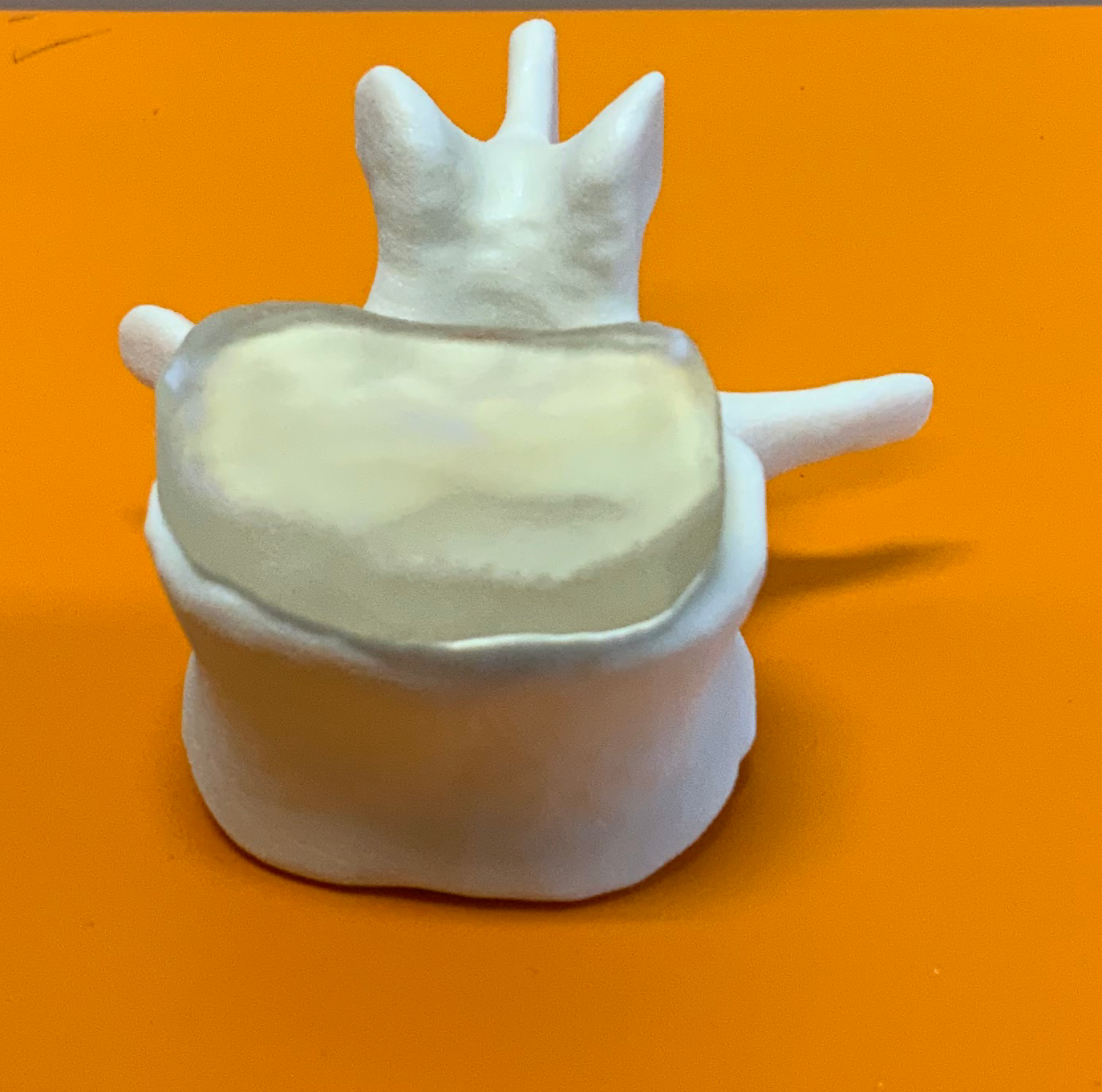}
	\end{tabular}
	\caption{\label{fig:3dprinting}Physical twin trough 3D printing}
\end{figure}
\section{Conclusions and future works}
By integrating computer vision and medical image processing approaches, we have introduced a comprehensive framework of an MS-DT to support personalised management of MSDs like preoperative planning and postoperative monitoring of the musculoskeletal  system, specifically for the vertebral spine, but general enough to be adapted to other anatomical districts. We have included a description of the different components and the double level of data integration performed in the connections module. The proposed MS-DT framework aims toward a personalised, multi modal data-driven healthcare approach. 

Leveraging the advancement in 3D printing, we also convert 3D segmented models of subparts of the anatomical district into patient-specific 3D subpart models with multi-material fabrication for accurate anatomy representation (Fig.~\ref{fig:3dprinting}). These models can assist in preoperative planning, surgical simulation, and medical training, improving outcomes with realistic, tactile replicas. 

All the integration methods included in the HME are general enough to support a great variety of imaging and sensors systems, however for clinical workflow integration the results of the integration could require a standardisation and exceptions management adjustments. 

In future work, we plan to include augmented reality visualisation modality to further help physicians and surgeons at various stages. Moreover we plan to integrate state of the art tools for biomechanical simulations to further improve the MS-DT's outcomes.

\section*{Acknowledgments}
All the authors are part of RAISE Innovation Ecosystem, funded by the European Union - NextGenerationEU and by the Ministry of University and Research (MUR), National Recovery and Resilience Plan (NRRP), Mission 4, Component 2, Investment 1.5, project “RAISE - Robotics and AI for Socio-economic Empowerment” (ECS00000035).

%Bibliography
\bibliographystyle{unsrt}  
\bibliography{MSDTRef}  
\end{document}